\documentclass[aps,prb,twocolumn]{revtex4-1}
\usepackage{amsmath}
\usepackage{amssymb}
\usepackage{natbib}
\usepackage{physics}
\usepackage{hyperref}
\usepackage{graphicx}
\usepackage{color}
\usepackage{ulem}

\begin{document}
\title{Charge and Spin Supercurrents in Magnetic Josephson Junctions with Spin Filters and Domain Walls}
\author{Samme M. Dahir, Anatoly F. Volkov, and Ilya M. Eremin}
\affiliation{Institut f\"ur Theoretische Physik III,
		Ruhr-Universit\"{a}t Bochum, D-44780 Bochum, Germany}

\begin{abstract}
We analyze theoretically the influence of domain walls (DWs) on the DC Josephson current in magnetic superconducting S\(_{m}\)/Fl/F/Fl/S\(_{m}\) junctions. The Josephson junction consists of two ''magnetic'' superconductors S\(_{m}\) (superconducting film covered by a thin ferromagnetic layer), spin filters Fl and a ferromagnetic layer F with or without DW (DWs). The spin filters Fl allow  electrons to pass with one specific spin orientation, such that the Josephson coupling is governed by a fully polarized long-range triplet component. In the absence of DW(s), the Josephson and spin currents are nonzero when the right and left filters, Fl\(_{r,l}\), pass electrons with equal spin orientation and differ only by a temperature-independent factor. They become zero when the \textbf{spins} of the triplet Cooper pairs passing through the Fl$_{r,l}$ have opposite directions. Furthermore, for the different chiralities of the injected triplet Cooper pairs the spontaneous currents arise in the junction yielding a diode effect. Once a DW is introduced, it reduces the critical Josephson current $I_{c}$ in the case of equal spin polarization and makes it finite in the case of opposite spin orientation. The critical current $I_{c}$ is maximal when the DW is in the center of the F film. A deviation of the DW from the center generates a force that pushes the DW to the center of the F film. In addition, we consider the case of an arbitrary number \(N\) of DW's, with the case \(N=2\) corresponding to a model system for a magnetic skyrmion.
\end{abstract}
	
\maketitle

Over the past decade, there has been a significant interest in studying the properties of superconductor/ferromagnet (S/F) hybrid
structures. One of the particular aspects of these heterostructures is related to remarkable phenomena caused by the magnetic interaction of topological textures in the superconductor (Abrikosov and Pearl vortices\cite{abrikosov_magnetic_1957,pearl_current_1964}) and in the ferromagnet (domain walls or skyrmions\cite{Panagopoulos_Emergence_2020,rosler_spontaneous_2006,1989thermodynamically,Soumyanarayanan_2016}). The interaction of vortices with the magnetic field in a ferromagnet may results in a spontaneous generation of vortices in the superconductor S in S/F bilayers \cite{lyuksyutov_magnetization_1998,Lyuksyutov,Milosevic2003,dahir_interaction_2019,Andriyakhina2021}. This effect occurs in the absence of a direct contact between the electron systems in S and F (no proximity effect) and is caused by the magnetic field generated by vortices or the magnetic textures. 

At the same time, the penetration of Cooper pairs into a ferromagnet (the proximity effect) leads to a number of further interesting effects. In particular,
the Josephson current in S/F/S junctions may change sign in a certain temperature interval (see
\cite{buzdin1982critical,buzdin1991josephson,ryazanov_coupling_2001,kontos_josephson_2002,sellier_temperature-induced_2003,Buzdin_quality_2006}
and also reviews \cite{golubov_current-phase_2004,buzdin_proximity_2005}. Another interesting effect is the triplet component which arises in S/F hybrid structures with a
inhomogeneous magnetisation $\mathbf{M(r)}$ in F. If the magnetization is uniform, the Cooper pairs penetrating into the ferromagnet consist of singlet and short-range triplet components, respectively. Both components penetrate into the ferromagnet over a short lengthscale \(\xi _{J}\approx \sqrt{D_{F}/J}\) (in the diffusive case), where \(D_{F}\) is the diffusion coefficient and \(J\) is the exchange field which, in most of ferromagnets, is much larger than the temperature \(T\) \cite{buzdin_proximity_2005,bergeret_odd_2005}.
If the magnetization \(\mathbf{M(r)}\) is non-homogeneous, as occurs, for example, in S/F\(_{m}\)/F structure, then a long range triplet component (LRTC) may occur in the system. Here, F\(_{m}\) is a weak ferromagnet with magnetization magnitude \(\mathbf{m}\) much less than \(\mathbf{M}\) and a direction is non-collinear to \(\mathbf{M}\). This component propagates into the F region over a long, compared to \(\xi _{J}\), length of the order of \(\xi _{T}\approx \sqrt{D_{F}/\pi T}\) \cite{bergeret_enhancement_2001,Kadigrobov_2001}. In this case, the superfluid component in most part of F
consists solely of triplet Cooper pairs. For example, the Josephson coupling
in S/F\(_{m}\)/F/F\(_{m}\)/S structure can be realized through the LRTC as it
was predicted \cite{Volkov_2003,Schoen_Half_2003,lofwander_interplay_2005,Houzet_long_range_2007,Volkov_Josephson_long_range_2007,Golubov_Josephson_odd_2007,Nazarov_triplet_2007} (see also reviews \cite{golubov_current-phase_2004,buzdin_proximity_2005,bergeret_odd_2005,eschrig2011spin,Houzet_Cryogenic_2019,balatsky_odd_2017} and
references therein) and observed experimentally \cite{keizer_spin_2006,sosnin_superconducting_2006,khaire_observation_2010,anwar_long_2012,salikhov_experimental_2009,robinson_enhanced_2010,kalenkov_triplet_2011,klose_optimization_2012,blamire_interface_2014,DiBernardo_signature_2015,massarotti_electrodynamics_2018,martinez_amplitude_2016,niedzielski_spin-valve_2018,caruso_tuning_2019,Norman_Spin_polarized_2020,Ahmad_electrodynamics_2020}. Interestingly, the long-range triplet Cooper pairs with spin-up and -down orientations penetrate the ferromagnet F regardless of the magnetization orientation \(\mathbf{M}\) \cite{Moor_nematic_2015} so that the spin current \(I_{sp}\) in S/F\(_{m}\)/F/F\(_{m}\)/S Josephson junctions is absent, whereas the charge current \(I_{Q}\) is non-zero. Only in the presence of spin filters at the S/F\(_{m}\) interfaces the current \(I_{sp}\) becomes finite.

In this manuscript, we calculate the Josephson charge \(I_{Q}\) and spin \(I_{sp}\) currents in the S\(_{m}\)/F\(l\)/F/Fl/S\(_{m}\) Josephson junctions under various conditions, where S\(_{m}=\)S/F\(_{m}\) is a conventional superconductor
covered by a thin ferromagnetic layer. First we consider the system without
DWs and calculate the currents \(I_{Q,sp}\) : a) in the absence or presence of
spin filters at the S/F\(_{m}\) interfaces, b) for equal or different polarizations or chiralities of the triplet Cooper pairs injected into F
from the left and right superconductors S. Most importantly, we also study the influence of
the domain walls in F (DWs) on the \(I_{Q}\) and \(I_{sp}\) in the dirty case when the condensate Green's functions \(\hat{f}\) obey
the Usadel equation. Within this approximation the Green's functions \(\hat{f}\) do not depend on the momentum direction \(\mathbf{p/|p|}\).
Therefore, according to the Pauli principle, the functions for the LRTC \(
\hat{f}(t,t^{\prime })\sim \langle c_{\uparrow }(t)c_{\uparrow }(t^{\prime
})\rangle \) are zero at coinciding times \(t=t^{\prime }\). In other
words, these are odd functions of the Matsubara frequency \(\omega \), \(\hat{f}
(\omega )=-\hat{f}(-\omega )\), so that summing over all \(\omega \) gives
zero: \(\hat{f}(t,t)\sim \sum_{\omega }\hat{f}(\omega )=0\).\ The triplet
odd-frequency Cooper pairs exist in any superconducting system if there is a
Zeeman interaction of electron spins and a magnetic or exchange field. This
case was studied long ago \cite{gorkov_ferromagnetism_1964,FFLO_1964,LOFF_1964,Kazumi_theory_1966,rusinov_theory_1969,Bulaevskii_coexistence_1985}. Unlike homogeneous superconductors with the Zeeman interaction, where the triplet component co-exists with the singlet one, the recently studied hybrid S/F systems allow the separation of triplet and singlet Cooper pairs.
In addition, we assume a weak proximity effect allowing linearization of the necessary equations
and the boundary conditions yielding simple analytical expressions for \(\hat{f}(r)\) and the currents \(I_{Q,sp}\).

Although the Josephson effect has been studied for similar structures in
various limiting cases (see references above as well as \cite{Eschrig_0pi_2008,Volkov_odd_2008,Manske_0pi_2009,Radoviclongrange2010,Haltermanspintronics2014,HaltermanJosephson2016}
), there is no systematic study of the dependence of the $I_{Q,sp}$ on spin
polarization, chiralities and the presence of the spin filters and DWs. In particular, we show that although the current $I_{Q}$ is zero for opposite polarization directions and different chiralities of injected Cooper pairs in the presence of spin filters, it becomes finite in the presence of DWs. We will consider an arbitrary number of DWs and pay a special attention
to the case of two DWs. The latter case may be regarded as a model of
magnetic texture such as skyrmion with $N=1$ winding number (like Bloch or Neel skyrmion) when the magnetisation profile $\mathbf{M}$ has
the same orientation outside the DWs and the opposite orientation between
DWs\cite{Panagopoulos_Emergence_2020,rosler_spontaneous_2006,1989thermodynamically,Soumyanarayanan_2016}.

\section{Basic Equations}

We consider an S\(_{m}\)/Fl/F/Fl/S\(_{m}\) Josephson junction with one or several
domain walls (DW) in the F film (wire). Schematically the considered system
is shown in Fig.1. The junction consists of two ''magnetic''
superconductors S\(_{m}\) and of two filters (Fl) which allow only electrons
with a single spin polarization, parallel or antiparallel to the z axis, to
pass through. The ''magnetic'' superconductors may be made of conventional
superconductors covered by ferromagnetic thin films with the magnetization
aligned parallel to the \(x\)- or \(y\)-axes. The magnetization vector \(\mathbf{	M}=(0,0,M)\) \ is supposed to be oriented along the \(z\)-axis. The filters may
be magnetic insulators selecting electrons with a spin collinear to the \(z\)
axis. The Cooper pairs penetrating into the F film due to proximity effect
consist  of triplet long-range components only. We assume that the proximity
effect is weak as it is the case in most of experimental setups. The Cooper pairs are described by a matrix quasiclassical Green’s function $\hat{f}(x)$, which is supposed to be small $|\hat{f}|<<1$. The function $\hat{f}(x)$ in the F film obeys the linearized Usadel equation\cite{golubov_current-phase_2004,buzdin_proximity_2005,bergeret_odd_2005,eschrig2011spin,balatsky_odd_2017}

\begin{widetext}

\begin{eqnarray}
	-\partial _{xx}^{2}\hat{f}+\kappa _{\omega }^{2}\hat{f}+\frac{i\kappa
	_{J}^{2}}{2}\left(n_{z}(x)\left[\mathit{\hat{X}}_{03},\hat{f}\right]_{+}+n_{k}(x)\left[\mathit{%
	\hat{X}}_{0k},\hat{f}\right]_{+}\right)=0\text{,}  \label{1}
\end{eqnarray}%
\end{widetext}
where \(\kappa _{\omega}^{2}=2|\omega |/D_{F}\), \(\kappa _{J}^{2}=2J$sgn$(\omega) /D_{F}\), \(J\) is a exchange field and \(D_{F}\) is the diffusion
coefficient in the F film which is assumed to be spin-independent. Note that
the quasiclassical equations with a spin-dependent \(D_{F}\) has been derived previously in various models\cite{Bergeret_Resitance_2002,Bobkova_gauge_2017}. The \(4\times 4\) matrix \(\mathit{\hat{X}}_{ik}(x)=\hat{\tau}_{i}\cdot \hat{\sigma}_{k}\) is a tensor product of the Pauli matrices in the Gor'kov-Nambu, \(\hat{\tau}_{i}\),
and spin space, \(\hat{\sigma}_{k}\), respectively. The square brackets are anticommutators
\([\mathit{\hat{X}}_{03},\hat{f}]_{+}=\mathit{\hat{X}}_{03}\cdot \hat{f}+\hat{f}\cdot \) \(\mathit{\hat{X}}_{03}\). The DW is assumed to be of the Bloch type
(\(k=y,z\)) and is described by a unit vector \(\mathbf{n}(x)=(0,n_{y}(x),n_{z}(x))\), where \(n_{y}=\sin \alpha (x)\), \(n_{z}(x)=\cos
\alpha (x)\). The angle \(\alpha (x)\) describes the DW profile: it is equal
to $0$ (left from DW) and to \(\pi \) (right from DW) far away
from DW. The characteristic size of the DW is \(d_{W}=\int dx\sin \alpha (x)\). In a general case Eq.(\ref{1}) can be solved only numerically. However, an exact solution can be also obtained under some assumptions like, for example, a piecewise
linear form of the DW \cite{bergeret_enhancement_2001,Aikebaier_DW_2020}. Here we will use a simple model assuming that the width of DW is small. Then, the last term in Eq.(\ref
{1}) can be written in the form \(\delta (x-l_{i})d_{DW}[\mathit{\hat{X}}_{02},\hat{f}]_{+}\), where \(l_{i}\) is a position of the DW. Then, the Usadel equation reduces to
\begin{equation}
	-\partial _{xx}^{2}\hat{f}+\kappa _{\omega }^{2}\hat{f}+i(\kappa _{J}^{2}/2)[%
	\mathit{\hat{X}}_{03},\hat{f}]_{+}=0\text{,}  \label{1b}
\end{equation}
with matching conditions at \(x=l_{i}\)
\begin{eqnarray}
	\hat{f}|_{l_{i}+0}-\hat{f}|_{l_{i}-0} &=&0\text{, }  \label{2} \\
	\partial _{x}\hat{f}|_{l_{i}+0}-\partial _{x}\hat{f}|_{l_{i}-0} &=&i(\kappa
	_{DW}/2)[\mathit{\hat{X}}_{02},\hat{f}]_{+}\text{,}  \label{2a}
\end{eqnarray}
where \(\kappa _{DW}=\kappa _{J}^{2}d_{DW}\).
\begin{figure}[t]
	\includegraphics[width=\columnwidth]{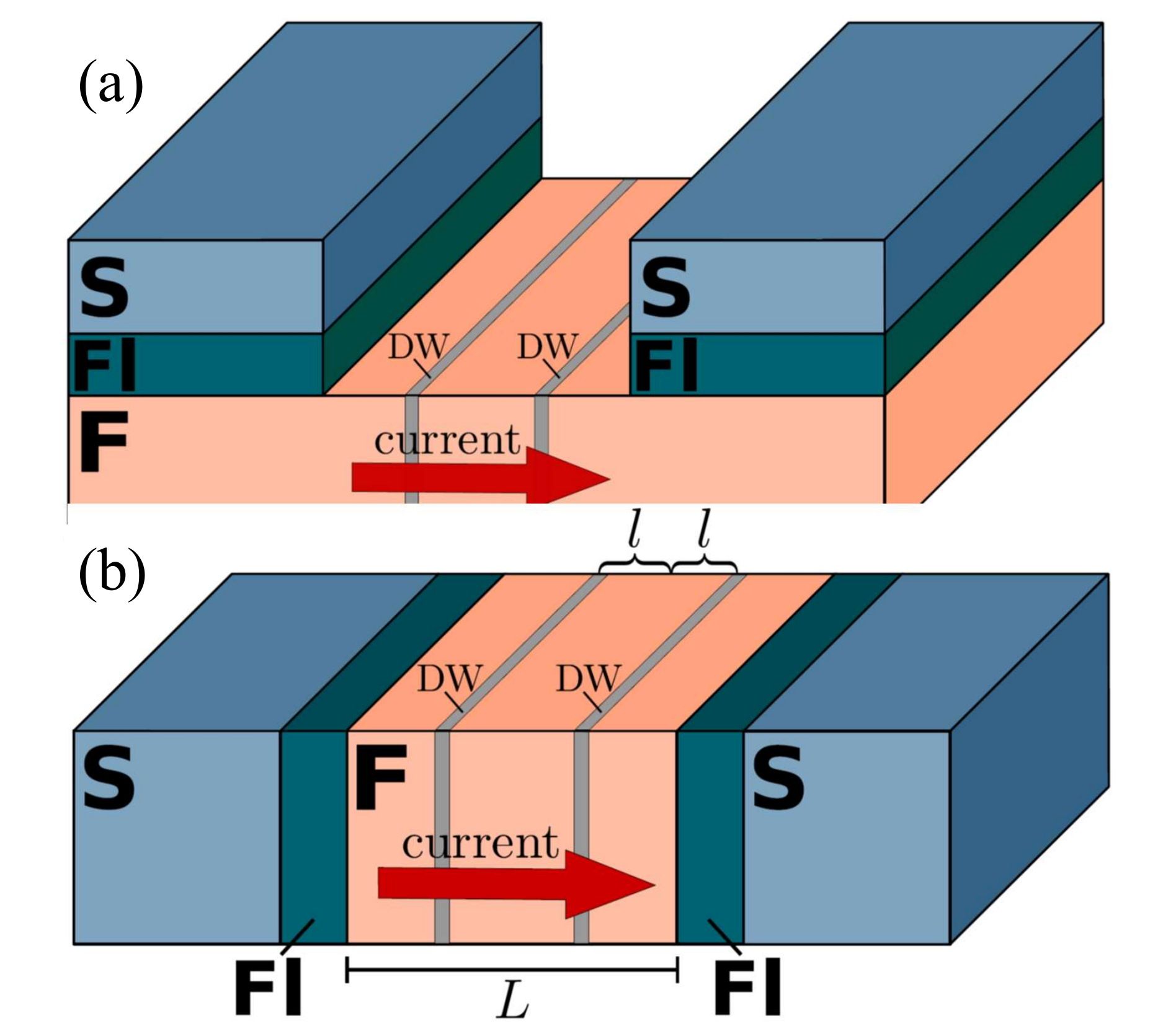} \vspace{-1mm}
	\vspace{-1mm}
	\caption{(Color online) Possible setups for the system under considerations shown examplarily for two domain walls.}
	\label{Fig1}
\end{figure}
%
%

A solution of Eq.(\ref{1b}) consists of a short-range and long-range
components, respectively. The first one decays on a short distance of the order of \(\xi
_{J}\approx \sqrt{D_{F}/J}\), while the second varies on a much longer
characteristic length of the order of \(\xi _{lr}\approx \sqrt{D_{F}/\pi T}\).
Observe that the condensate matrix Green's function \(\hat{f}\) is
off-diagonal in the Gor'kov-Nambu space, {\it i. e.}  \(\hat{f}\) is proportional to \( \hat{\tau}
_{1},\hat{\tau}_{2}\) matrices. In addition, the long-range triplet component
(LRTC), \(\hat{f}_{lr}\), is also off-diagonal in the spin space, {\it i. e.}  \(\hat{f}\) is proportional \(
\hat{\sigma}_{1},\hat{\sigma}_{2}\) matrices such that the third term in Eq.(\ref{1b})
for this component vanishes. This means that in a general case the matrix
LRTC \(\hat{f}_{lr}\) obeying Eq.(\ref{1b}) can be written in the form
\begin{equation}
	\hat{f}_{lr}(x)=\sum_{\{i,k\}}a_{ik}(x)\mathit{\hat{X}}_{ik}\text{,}
	\label{2b}
\end{equation}
where \(\{i;k\}=\{1,2;1,2\}\). A concrete form of the LRTC is determined by
the boundary conditions at \(x=\pm L\). These boundary conditions, originally
employed in Refs.\cite{Volkov_transport_2012,Eschrig_boundary_2015,Bergeret_anomalous_2017,zaitsev_negative_2018}, can be represented in a simple form
\begin{eqnarray}
	\partial _{x}\hat{f}|_{x=\pm L} &=&\pm \kappa _{b}\mathit{\hat{X}}_{r,l}F_{S-}\text{, }  \label{3} \\
	F_{S-} &=&i\Delta \text{Im}(1/\zeta _{\omega +})  \label{3a}
\end{eqnarray}%
where \(\mathit{\hat{X}}_{r,l}=\hat{T}\cdot \mathit{\hat{X}}_{m}\cdot \hat{T}^{\dagger }\), \(\kappa _{b}=1/(R_{b}\sigma _{F})\), \(R_{b}\) is an interface (barrier) resistance per unit area, \(\sigma _{F}\) is conductivity
of the F film, \(\zeta _{\omega \pm }=\sqrt{(\omega \pm iJ_{m})^{2}+\Delta^{2}}\), \(J_{m}\) is an exchange field in the F\(_{m}\). The functions \(\mathit{\hat{X}}_{r,l}F_{S-}=(\hat{T}\cdot \mathit{\hat{X}}_{m}\cdot \hat{T}^{\dagger })F_{S-}\) are the Green's functions of the Cooper pairs passing through the right (left) filters. The matrix coefficient \(\hat{T}\) describes the tunneling of Cooper pairs through the filters and is defined as \cite{Volkov_transport_2012}
\begin{equation}
	\hat{T}=(\mathcal{T}+\mathcal{U}\mathit{\hat{X}}_{33})/\sqrt{2}  \label{3b}
\end{equation}
The term \(\mathit{\hat{X}}_{m}F_{S-}\equiv \hat{f}_{m}\) in Eq.(\ref{3a}) is
a matrix Green's function in the weak ferromagnet F\(_{m}\) where the
function \(F_{S-}\) is an odd function of the Matsubara frequencies \(\omega
=\pi T(2n+1)\) and describes the triplet component. The form of the matrix \(
\mathit{\hat{X}}_{m}(\pm L)\equiv \) \(\mathit{\hat{X}}_{m}|_{l,r}\) depends on
the chirality of the LRTC, i.e., on the orientation of the magnetization \(
\mathbf{m}\) in the F\(_{m}\) ferromagnetic film \cite{Moor_chirality_2015}
\begin{eqnarray}
	\mathit{\hat{X}}_{m}^{\text{(x)}} &=&\mathit{\hat{X}}_{11}\text{, }\mathbf{m}%
	||\mathbf{e}_{x}  \label{D1} \\
	\mathit{\hat{X}}_{m}^{\text{(y)}} &=&\mathit{\hat{X}}_{12}\text{, }\mathbf{m}%
	||\mathbf{e}_{y}  \label{D1a}
\end{eqnarray}
The matrices \(\mathit{\hat{X}}_{m}^{\text{(x)}}\), \(\mathit{\hat{X}}_{m}^{\text{(y)}}\) describe triplet Cooper pairs with spin up and down, which have different chirality. The filter action converts the \(\{\mathit{\hat{X}}_{m}^{\text{(x)}},\mathit{\hat{X}}_{m}^{\text{(y)}}\}\) matrices to  \(\{\mathit{\hat{X}}_{l,r}^{\text{(x)}},\mathit{\hat{X}}_{l,r}^{\text{(y)}}\}\),
where
\begin{eqnarray}
	\mathit{\hat{X}}_{l,r}^{\text{(x)}} &=&\mathit{\hat{X}}_{11}-s_{l,r}\mathit{%
	\hat{X}}_{22}\text{, }\mathbf{m}||\mathbf{e}_{x}  \label{D2} \\
	\mathit{\hat{X}}_{l,r}^{\text{(y)}} &=&\mathit{\hat{X}}_{12}+s_{l,r}\mathit{%
	\hat{X}}_{21}\text{, }\mathbf{m}||\mathbf{e}_{y}  \label{D2a}
\end{eqnarray}
The parameter \(s=2\)Re\((\mathcal{TU}^{\ast })/(|\mathcal{T}|^{2}+|\mathcal{U}|^{2})\) characterizes the degree of spin-up and spin-down polarization of the triplet Cooper pairs injected into the film F. If \(\mathcal{U}=0\), Cooper pairs with up and down spins penetrate into the F film with equal probabilities, and therefore the number of the triplet pairs with both spin orientations in the F is the same. This case has been called nematic\ LRTC in Ref.\cite{Moor_nematic_2015}. If \(|\mathcal{T}|=|\mathcal{U}|=1\), then \(s=\pm 1\), and the triplet Cooper pairs are fully polarized with total spin parallel or antiparallel to the \(z\)- axis. Note that a magnetic half-metal can be used as a spin filter. The case of \(s=0\) corresponds to the absence of filters at the F\(_{m}\)/F interfaces.

Eqs.(\ref{3a}-\ref{D2}) are a generalization of the Kupriyanov-Lukichev
boundary conditions \cite{Kurpianov1988} which in turn were obtained from
the Zaitsev's boundary conditions \cite{zaitsev_quasiclassical_1984} (see also Ref\cite{Volkov_BC_1997}, where the applicability of the Kupriyanov-Lukichev boundary conditions is discussed).

Till now we assumed that the phases of the order parameter in the superconductors S are chosen equal to zero. The presence of the phases \(\pm \varphi /2\) at S\(_{r,l}\) can be easily introduced via a gauge transformation \(\hat{S}_{\varphi}=\exp (\pm i\mathit{\hat{X}}_{30}\varphi /4)\): \(\hat{g}_{S,\varphi }=\hat{S}_{\varphi }\cdot \) \(\hat{g}_{S}\cdot \hat{S}_{\varphi }^{\dagger}\) (see,
for example, \cite{Zaitsev_Theory_1979}) so that the boundary condition (\ref{3a}) can be
written as
\begin{equation}
	\partial _{x}\hat{f}|_{\text{ }x=\pm L}=\pm \kappa _{b}[\cos (\varphi /2)\pm
	i\mathit{\hat{X}}_{30}\sin (\varphi /2)]\mathit{\hat{X}}_{l,r}F_{S-}.
	\label{D3}
\end{equation}
The matrix condensate function \(\mathit{\hat{X}}_{m}^{\text{(x,y)}}F_{S-}\) describes a short-range triplet component in the film F\(_{m}\), but it
becomes a long-range one in the F film because of the non-collinearity of the the magnetization	vectors \(\mathbf{m}\) and \(\mathbf{M}\). Note that the
functions \(\mathit{\hat{X}}_{12}F_{S-}\) and \(\mathit{\hat{X}}_{21}F_{S-}\), written explicitly, consist of triplet components with up and down spins \(\mathit{\hat{X}}
_{12}F_{S-}\sim \langle c_{\uparrow }(t)c_{\uparrow }(0)\rangle +\langle c_{\downarrow }(t)c_{\downarrow }(0)\rangle \), \(\mathit{\hat{X}}_{21}F_{S-}\sim \langle c_{\uparrow }(t)c_{\uparrow }(0)\rangle -\langle c_{\downarrow }(t)c_{\downarrow }(0)\rangle \) so that the function \(\mathit{\hat{X}F}_{S-}=(\mathit{\hat{X}}_{12}\pm \mathit{\hat{X}}_{21})F_{S-}\) describes the Cooper pairs polarized in one direction, see Appendix A for further details.

Knowing the Green's functions \(\hat{f}\), we can readily calculate the charge \(I_{Q}=\mathbf{I}_{Q}\mathbf{\cdot e}_{x}\) and the spin currents \(I_{sp}=\mathbf{I}_{sp}^{(z)}\cdot \mathbf{e}_{x}\) using the following expressions 
\begin{eqnarray}
	I_{Q} &=&\frac{\sigma _{F}}{e}2\pi T\sum_{\omega \geqslant 0}I_{Q,\omega }%
	\text{,}  \label{6} \\
	I_{sp} &=&\mu _{B}\frac{\sigma _{F}}{e^{2}}2\pi T\sum_{\omega \geqslant
		0}I_{sp,\omega }  \label{6a}
\end{eqnarray}
where the ''spectral'' currents \(I_{Q,\omega}\) and \(I_{sp,\omega}\) are
defined as
\begin{eqnarray}
	I_{Q,\omega } &=&(i/4)\text{Tr}\{(\hat{\tau}_{3}\cdot \hat{\sigma}_{0})\hat{f%
	}_{LR,0}\partial _{x}\hat{f}\}\equiv i\{\hat{f}\partial _{x}\hat{f}\}_{30}
	\label{7} \\
	I_{sp,\omega } &=&(i/4)\text{Tr}\{(\hat{\tau}_{0}\cdot \hat{\sigma}_{3})\hat{%
		f}\partial _{x}\hat{f}\}\equiv i\{\hat{f}\partial _{x}\hat{f}\}_{03}
	\label{7a}
\end{eqnarray}
and further details are given in Appendix B.
Similar formulas were used in \cite{Moor_chirality_2015,Aikebaier_DW_2020,yokoyama2021anisotropic}. Observe that the traces in the Nambu space for charge and spin currents are actually different, which was often overlooked previously.

In order to find the Josephson current, we need to solve Eq.(\ref{1b})\ with
the matching conditions (\ref{2}) and boundary conditions (\ref{3a}).

We first consider the case of the F film with a uniform magnetization, $\mathbf{M}=(0,0,M)$, without DWs. Although such magnetic Josephson junctions have been already studied previously in different limiting cases (ballistic
and diffusive) using various mostly numerical techniques \cite{Volkov_2003,Schoen_Half_2003,lofwander_interplay_2005,Houzet_long_range_2007,Volkov_Josephson_long_range_2007,Golubov_Josephson_odd_2007,Nazarov_triplet_2007,Eschrig_0pi_2008,Volkov_odd_2008,Manske_0pi_2009,Radoviclongrange2010,Haltermanspintronics2014,HaltermanJosephson2016,Haltermantransport2018,Bergeretcharge2019}
we will discuss the main results in the dirty case and in the limit of the weak proximity effect. Then the formulas for currents acquire a simple analytical form, not known previously, that allows for a straightforward physical interpretation.  In addition, we will focus our study on the case of fully polarized triplet Cooper pairs of different chiralities. 

In particular, the solution of Eq.(\ref{1b}), \(\hat{f}_{LR,0}\), which obeys the boundary conditions (\ref{3a}) has the form
\begin{equation}
	\hat{f}_{lr,0}=\hat{C}\frac{\cosh (\kappa _{\omega }x)}{\sinh (L\kappa
		_{\omega })}+\hat{S}\frac{\sinh (\kappa _{\omega }x)}{\cosh (L\kappa
		_{\omega })}  \label{4}
\end{equation}
with
\begin{eqnarray}
	\hat{C} &=&\frac{\kappa _{b}}{2\kappa _{\omega }}\left[\mathit{\hat{X}}_{+}\cos\left(\frac{\varphi }{2}\right)+i\mathit{\hat{X}}_{30}\cdot \mathit{\hat{X}}_{-}\sin\left(\frac{\varphi }{2}\right)\right]F_{S-}  \label{5a} \\
	\hat{S} &=&\frac{\kappa _{b}}{2\kappa _{\omega }}\left[\mathit{\hat{X}}_{-}\cos\left(\frac{\varphi}
	{2}\right)+i\mathit{\hat{X}}_{30}\cdot\mathit{\hat{X}}_{+}\sin\left(\frac{\varphi}{2}\right)\right]F_{S-}  \label{5b}
\end{eqnarray}
where we have defined \mbox{\(\mathit{\hat{X}}_{\pm}=\mathit{\hat{X}}_{r}\pm\mathit{\hat{X}}_{l}\)}. Substituting \(\hat{f}_{LR,0}\) from Eq.(\ref{4}) into Eqs.(\ref{7}-\ref{7a}), we obtain
\begin{eqnarray}
	I_{Q,\omega } &=&\tilde{I}_{\omega}\left[-i\{\mathit{\hat{X}}_{r}\cdot \mathit{\hat{X}}%
	_{l}\}_{30}\cos \varphi +\{\mathit{\hat{X}}_{r}\cdot \mathit{\hat{X}}%
	_{l}\}_{00}\sin \varphi \right]  \label{8} \\
	I_{sp,\omega } &=&\tilde{I}_{\omega}\left[-i\{\mathit{\hat{X}}_{r}\cdot \mathit{\hat{X}}%
	_{l}\}_{03}\cos \varphi +\{\mathit{\hat{X}}_{r}\cdot \mathit{\hat{X}}%
	_{l}\}_{33}\sin \varphi \right] \label{8a} \\
	\tilde{I}_{\omega } &=&\frac{(\kappa _{b}F_{S-})^{2}}{\kappa _{\omega
		}\sinh (2L\kappa _{\omega })}\text{.}  \label{8b}
\end{eqnarray}
and \(\kappa _{\omega }=\sqrt{2|\omega |/D_{F}}\). Observe that the
matrices \(\hat{C},\hat{S}\) and \(\mathit{\hat{X}}_{r},\mathit{\hat{X}}_{l}\)
anticommute with matrices \(\mathit{\hat{X}}_{30},\mathit{\hat{X}}_{03}\) so
that the traces \(\{\hat{C}^{2}\}_{30},\{\hat{C}^{2}\}_{03}\) etc. are equal to zero. In the following we calculate the charge and spin currents for different cases in detail. 
\subsection{Currents in the absence of filters.}

For the case of equal chiralities of the triplet Cooper pairs injected from the right (left) S/Fl interfaces (\(\mathbf{m}_{l}||\mathbf{m}_{r}||\mathbf{e}_{x}\) or \(\mathbf{m}_{l}||\mathbf{m}_{r}||\mathbf{e}_{y}\)) and defining \(\mathit{\hat{X}}_{r}=\mathit{\hat{X}}_{l}=\mathit{\hat{X}}_{11}\equiv \mathit{\hat{X}}_{m}^{\text{(x)}}\) or \(\mathit{\hat{X}}_{r}=\mathit{\hat{X}}_{l}=\mathit{\hat{X}}_{12}\equiv \mathit{\hat{X}}_{m}^{\text{(y)}}\), the charge and spin ''spectral'' currents are
\begin{eqnarray}
	I_{Q,\omega }^{\text{(x,x)}} &=&I_{Q,\omega }^{\text{(y,y)}}=\tilde{I}%
	_{\omega }\sin \varphi  \label{D4} \\
	I_{sp,\omega }^{\text{(x,x)}} &=&I_{sp,\omega }^{\text{(y,y)}}=0  \label{D4a}
\end{eqnarray}
{\it i.e.} the charge current has the usual form \(I_{Q,\omega }=\tilde{I}\sin \varphi \) whereas the spin current is zero. For the case of different chiralities (\(\mathit{\hat{X}}_{r}=\mathit{\hat{X}}_{12}\), \(\mathit{\hat{X}}_{l}=\mathit{\hat{X}}_{11}\)) the currents are given by
\begin{eqnarray}
	I_{Q,\omega }^{\text{(x,y)}} &=&0  \label{D5} \\
	I_{sp,\omega }^{\text{(x,y)}} &=&\tilde{I}_{\omega }\cos \varphi  \label{D5a}
\end{eqnarray}
where indices (x,x) and (x,y) refer to the chirality of the Cooper pairs penetrating the film F on the right and on the left that is, \(I_{Q,\omega }^{\text{(x,y)}}\sim \{\mathit{\hat{X}}_{r}^{\text{(x)}}\cdot \mathit{\hat{X}}_{l}^{\text{(y)}}\}\). We also note an important feature of the obtained currents. In particular, the critical ''spectral'' current \(\tilde{I}_{\omega }\) in the considered S\(_{m}\)/Fl/F/Fl/S\(_{m}\) junction has the sign opposite to that in S/N/S Josephson
junction since in the latter case the critical current \(\tilde{I}_{S/N/S}\sim F_{S}^{2}>0\), while in the system under consideration \(\tilde{I}_{\omega }\sim F_{S-}^{2}<0\) (see Eq.(\ref{3a})), here \(F_{S}=\Delta /\sqrt{\omega ^{2}+\Delta ^{2}}\). This is a simple representation of the fact that the LRTC leads to a \(\pi \)-Josephson coupling.

Observe that the Josephson current \(I_{Q,\omega }^{\text{(x,x)}}\) is finite for collinear orientations of the magnetic moments \(\mathbf{m}\) in the left
and right films F\(_{m}\) and is zero (\(I_{Q,\omega }^{\text{(x,y)}}=0\)) for the orthogonal orientations of the vectors \(\mathbf{m}_{l,r}\). The opposite is true for the spin current. It is zero in the case of vectors \(\mathbf{m}_{l}=\mathbf{m}_{r} \) and is finite if \(\mathbf{m}_{l}\cdot \mathbf{m}_{r}=0\), i.e.,
when the vectors \(\mathbf{m}_{r,l}\) are orthogonal. Moreover, in the second case a spontaneous spin current arises in the system even when the phase
difference \(\varphi\) is zero.

The formulas for the currents (\ref{D4}-\ref{D5a}) are derived for the case when the vectors \(\mathbf{m}_{r,l}\) lie in the plane perpendicular to the \(z\)-axis so that \(\mathbf{m}_{r,l}\cdot \mathbf{e}_{z}=\cos \alpha_{r,l}=0\). They can be easily generalized for the arbitrary angles \(\alpha_{r,l}\). Taking into account that only the components \(\mathbf{m}_{r,l}\cdot\mathbf{e}_{x,y}\) contribute to the LRTC, in a more general case the currents \(\tilde{I}\) are equal to
\begin{equation}
	\tilde{I}_{\alpha }=\tilde{I}\sin \alpha _{r}\sin \alpha _{l}  \label{D6}
\end{equation}
The formulas for the charge and spin currents \(I_{Q,sp,\omega}\) are represented in Table 1. The angles \(\alpha _{r,l}\) are chosen to be equal to
\(\pi /2\) so that \(\sin \alpha _{r}=\sin \alpha _{l}=1\).

\begin{widetext}

\begin{table}[t!]
\caption{Summary of the charge and spin supercurrents in magnetic Josephson junctions with spin filters and their modifications due to domain walls for different ferromagnetic filter orientations.}
\centering
\begin{tabular}{c}
\includegraphics[width=\textwidth]{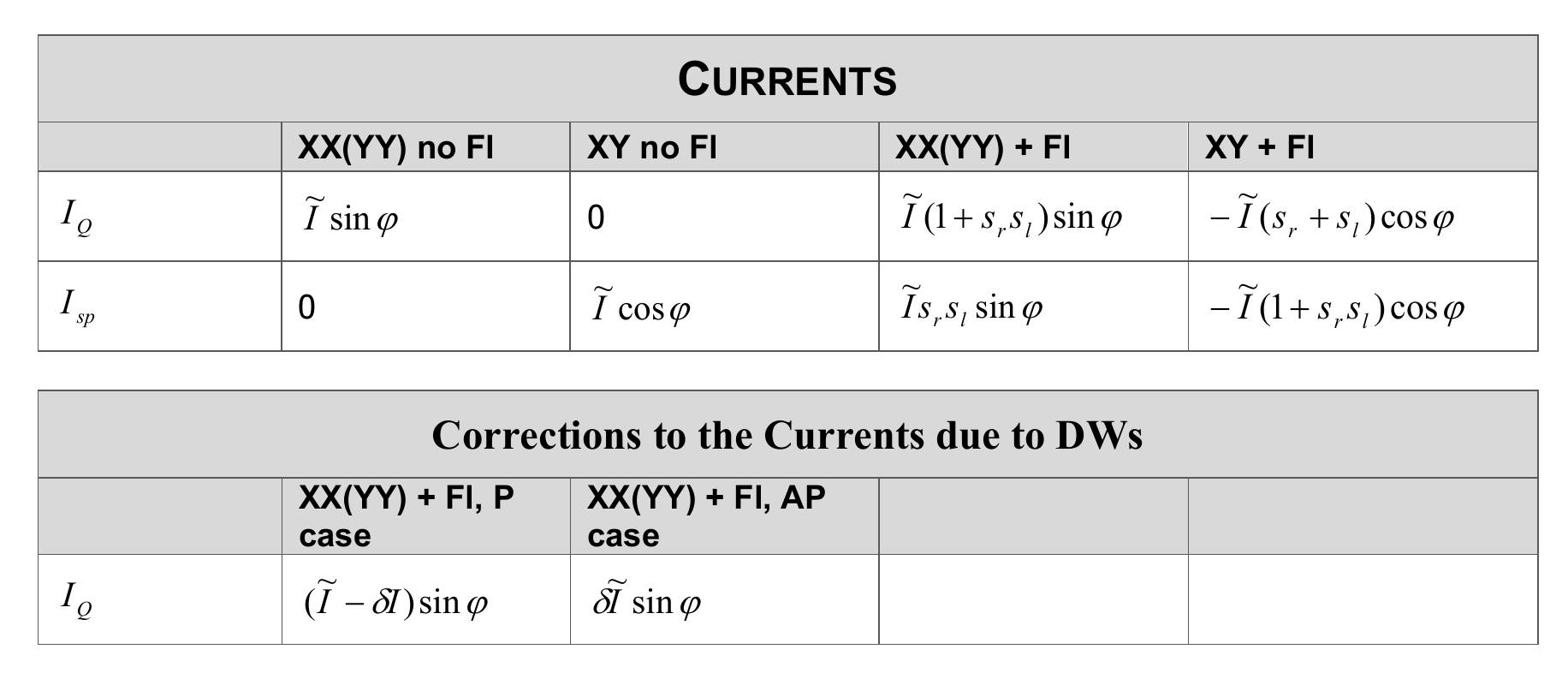} 
\end{tabular}
  \label{tab:1}
\end{table}

\end{widetext}

\subsection{Currents in the presence of filters.}

Now we calculate the currents for the case of a uniform of \(\mathbf{M}\) in F and in the presence of filters at the interfaces F/F\(_{m}\). We remind that in the absence of spin filters, the currents
are spin independent. As we show below the presence of spin filters makes both currents spin dependent. For the case of equal chiralities, i.e., \(\mathit{\hat{X}}_{r}=\mathit{\hat{X}}_{l}=\mathit{\hat{X}}^{\text{(}\nu \text{)}}\) (\(\nu =x\) or \(y\)), the charge and spin currents can be found by using formulas for \(\mathit{\hat{X}}^{\text{(x)}}\) and \(\mathit{\hat{X}}^{\text{(y)}}\), Eq.(\ref{D1}),
\begin{eqnarray}
	I_{Q,\omega }^{\text{(x,x)}} &=&I_{Q,\omega }^{\text{(y,y)}}=\tilde{I}_{\omega}
	(1+s_{r}s_{l})\sin \varphi \text{,}  \label{9} \\
	I_{sp,\omega }^{\text{(x,x)}} &=&I_{sp,\omega }^{\text{(y,y)}}=\tilde{I}_{\omega}
	(s_{r}+s_{l})\sin \varphi  \label{9a}
\end{eqnarray}
These formulas show that for parallel spin orientations of fully polarized triplet Cooper pairs injected from the right and left superconductors, the
values of the coefficients \(1+s_{r}s_{l}=2\,\)\ and \(s_{r}+s_{l}=2\) sgn(\(s\)) are the same, but the direction of the spin current depends on the sign of \(s\). In the case of opposite spin polarization both currents are zero.

For different chiralities \(\mathit{\hat{X}}_{r}=\mathit{\hat{X}}^{\text{(x)}} \), \(\mathit{\hat{X}}_{l}=\mathit{\hat{X}}^{\text{(y)}}\) we find
\begin{eqnarray}
	I_{Q,\omega }^{\text{(x,y)}} &=&\tilde{I}(s_{r}+s_{l})\cos \varphi \text{,}
	\label{10} \\
	I_{sp,\omega }^{\text{(x,y)}} &=&\tilde{I}(1+s_{r}s_{l})\cos \varphi
	\label{10a}
\end{eqnarray}
With equal spin polarizations (\(s_{r}=s_{l}\))  (\(s_{r}=s_{l}\)), the spontaneous charge and spin currents occur in this case even in the absence of a phase difference.
Interestingly, the direction of the spontaneous charge current depends on the sign of spins $s$ of injected triplet Cooper pairs. In the case of
opposite spin polarization, these currents disappear. Note that the spontaneous currents may lead to the Josephson diode eﬀect, see Ref.\onlinecite{Pal_2021} and references therein. The conclusion about the possibility of spontaneous currents in different models of superconducting magnetic systems (with or without spin-orbit interaction) have been obtained earlier \cite{Nazarov_triplet_2007,Buzdin_direct_2008,Moor_nematic_2015,Bergeret_anomalous_2017,Buzdin_sponatneous_2017}
(see also recent papers \cite{Buzdin_phase_transition_2021,montiel_spin_2021} and references therein).  For convenience we summarize the results for the charge and spin currents in Table 1.

\section{Modifications of the Currents due to DWs.}

Next we consider the modifications of the currents, obtained above, for the case of the domain wall in the F film.
We restrict our analysis to the case of equal chiralities (the generalization to the case of different chiralities is straightforward) and also assume that the spacing between the nearest DWs is much larger than the decay length of the short-range component \(\hat{f}_{sr}\), i.e., \(|l_{1}-l_{2}|\gg \xi _{J}\approx \sqrt{D_{F}/J}\). The main effect of the domain wall is  the creation of a short-range triplet component,  which results in a correction \(\delta \hat{f}_{lr}\) to the long-range component \(\hat{f}_{lr,0}\) defined by Eq.(\ref{4}). While the short-range component exists only near each DW, the LRTC extends over a larger distance, which can be of the order of \(L\). In particular, 
the correction \(\delta \hat{f}_{lr}\) arises due to
matching conditions for the function \(\delta \hat{f}_{lr}(x)\) at \(x=l_{i}\), where \(l_{i}\) is the coordinate of a DW. These conditions for \(\delta \hat{f}_{lr,0}(x)\) and its partial derivative are
\begin{eqnarray}
	\lbrack \delta \hat{f}_{lr}]|_{l_{i}} &=&0\text{, }  \label{DW1} \\
	\lbrack \partial _{x}\delta \hat{f}_{lr}]|_{l_{i}} &=&i(\kappa _{DW}/2)[%
	\mathit{\hat{X}}_{02},\hat{f}_{sr}(l)]_{+}\text{,}  \label{DW1a}
\end{eqnarray}
As usual, these are complemented \ by the boundary conditions
\begin{equation}
	\partial _{x}\delta \hat{f}_{lr}|_{\pm L}=0\text{, }  \label{DW2}
\end{equation}
In the presence of several DWs, the solution for \(\delta \hat{f}_{lr}\) can be represented in the form
\begin{equation}
	\delta \hat{f}_{lr}(x)=\sum_{i}\delta \hat{f}_{lr}^{(i)}(x)\text{, }
	\label{DW3}
\end{equation}
where the \(\delta \hat{f}_{lr}^{(i)}(x)\) is a perturbation of the LRTC generated by the \(i-\)th DW. In order to find this function, one needs to
determine a short-range component \(\hat{f}_{sr}^{(i)}\) produced by the \(i-\)th DW, which we do in the next subsection.

\subsection{Short-range Component generated by the Domain Wall}

The short-range component obeys Eq.(\ref{1b}) and matching conditions (\ref{2}-\ref{2a}) that can be written as
\begin{eqnarray}
	\lbrack \hat{f}_{sr}]|_{l} &=&0\text{, }  \label{Sr1} \\
	\lbrack \partial _{x}\hat{f}_{sr}]|_{l} &=&i\frac{\kappa _{DW}}{2}[\mathit{%
		\hat{X}}_{02},\hat{f}_{lr,0}]_{+}\text{,}  \label{Sr1a}
\end{eqnarray}%
where we also dropped the subindex \(i\) in \(l_{i}\) for simplicity. Taking into account Eqs.(\ref{2a},%
\ref{5a}-\ref{5b}), we can rewrite Eq.(\ref{Sr1a}) as follows
\begin{equation}
	\lbrack \partial _{x}\hat{f}_{sr}]|_{l}=i\kappa _{DW}\left(\hat{C}_{2}\frac{\cosh
		\tilde{l}}{\sinh \tilde{L}}+\hat{S}_{2}\frac{\sinh \tilde{l}}{\cosh \tilde{L}%
	}\right)  \label{Sr2a}
\end{equation}
where \(\tilde{l}=\kappa _{\omega }l\), \(\tilde{L}=\kappa _{\omega }L\) and \(
\hat{C}_{2}=[\mathit{\hat{X}}_{02},\hat{C}]_{+}\), \(\hat{S}_{2}=[\mathit{\hat{X}}_{02},\hat{S}]_{+}\). A solution for the short-range component, Eq.(\ref{1b}), obeying the matching conditions (\ref{Sr1},\ref{Sr1a}) in the vicinity of \(i-\)th DW can written in the form 
\begin{equation}
	\hat{f}_{sr}=\hat{f}_{sr}^{(\text{A})}\cos (\varphi /2)+\hat{f}_{sr}^{(\text{B})}\sin (\varphi /2)\text{.}  \label{Sr3}
\end{equation}
where the matrices \(\hat{f}_{sr}^{(\text{A,B})}\) Green's functions contain exponentially decaying functions
\begin{widetext}

\begin{equation}
	\hat{f}_{sr}^{\text{(A)}}(x)=-i{\Big \{}%
	\begin{array}{c}
		A_{+}\mathit{\hat{X}}_{n+}\exp (K_{+}(x-l))+A_{-}\mathit{\hat{X}}_{n-}\exp (K_{-}(x-l))\text{, }x<l \\
		\bar{A}_{+}\mathit{\hat{X}}_{n+}\exp (-\bar{K}%
		_{+}(x-l))+\bar{A}_{-}\mathit{\hat{X}}_{n-}\exp (-%
		\bar{K}_{-}(x-l))\text{, }l<x\text{,}%
	\end{array}%
	\;.  \label{Sr4}
\end{equation}
\end{widetext}
where \(K_{\pm }^{2}=\kappa _{\omega }^{2}\pm i\kappa _{J}^{2}\) and \(n=1,2\)
for \(y\), \(x\)- chiralities, \(\bar{K}_{\pm }=K_{\mp }\) and \mbox{\(\mathit{\hat{X}}_{n\pm}=(\mathit{\hat{X}}_{n0}-\mathit{\hat{X}}_{n3})\)}. The matrix \(\hat{f}_{sr}^{(\text{B})}\) is equal
\begin{equation}
	\hat{f}_{sr}^{\text{(B)}}(x)=i\mathit{\hat{X}}_{30}\cdot \hat{f}_{sr}^{\text{%
			(A)}}(x)\text{,}  \label{Sr4a}
\end{equation}
with the replacement \(\ A\Rightarrow B\). The matching condition (\ref{Sr1}) yields
\begin{eqnarray}
	A_{+} &=&A_{-}=\bar{A}_{-}=\bar{A}_{+}\equiv A\text{,}  \label{Sr5a} \\
	B_{+} &=&B_{-}=\bar{B}_{-}=\bar{B}_{+}\equiv B  \label{Sr5b}
\end{eqnarray}
The coefficients \(A\) and \(B\) are determined from Eq.(\ref{Sr1a}). In what follows we consider several cases. \\
{\it (a) The \(x\)- chirality, parallel \(s_{r,l}\) orientations.} In this case, \(\mathit{\hat{X}}_{r}\mathit{=\hat{X}}_{l}=(\mathit{	\hat{X}}_{11}-s\mathit{\hat{X}}_{22})\). The coefficients \(A_{P}^{(x)}\), \(B_{P}^{(x)}\) are equal to
\begin{eqnarray}
	A_{P}^{(x)} &=&-sr_{\omega }\frac{\cosh \tilde{l}}{\sinh \tilde{L}}F_{S-}%
	\text{\ }  \label{Sr6} \\
	B_{P}^{(x)} &=&-sr_{\omega }\frac{\sinh \tilde{l}}{\cosh \tilde{L}}F_{S-}
	\label{Sr6a}
\end{eqnarray}
where \(r_{\omega }=\) (\(\kappa _{DW}\kappa _{b})/(K_{0}\kappa _{\omega })\) and \(K_{0}=4\)Re\(K_{+}\).\\

{\it b) The \(x\)- chirality, antiparallel spin orientations, i. e., \(s_{r}=s=-s_{l}\).}
In this case  \(\mathit{\hat{X}}_{r,l}\mathit{=}(\mathit{\hat{X}}_{11}\mp s\mathit{\hat{X}}_{22})\). The
coefficients \(A_{AP}^{(x)}\), \(B_{AP}^{(x)}\) are given by: \(A_{AP}^{(x)}=\) \(
B_{P}^{(x)}\), \(B_{AP}^{(x)}=A_{P}^{(x)}\).\\

{\it c) The \(y\)- chirality, parallel (antiparallel) \(s_{r,l}\) orientations.}
Then, \(\mathit{\hat{X}}_{l,r}\mathit{=}(\mathit{\hat{X}}_{12}-s_{l,r}\mathit{\hat{X}}_{21})\) and the coefficients \(A^{(y)}\), \(B^{(y)}\) are equal to
\begin{eqnarray}
	A_{P}^{(y)} &=&A_{AP}^{(y)}=A_{P}^{(x)}/s\text{,}  \label{Sr7} \\
	B_{P}^{(y)} &=&B_{AP}^{(y)}=B_{P}^{(x)}/s.  \label{Sr7a}
\end{eqnarray}
In the next section, we calculate the function \(\delta \hat{f}_{lr,0}(x)\).

\subsection{Correction to the LRTC due to a domain wall}

Finally, the correction \(\delta \hat{f}_{lr}(x)\) obeys the equation
\begin{equation}
	-\partial _{xx}^{2}\delta \hat{f}_{lr}+\kappa _{\omega }^{2}\delta \hat{f}%
	_{lr}=0\text{,}  \label{LR0}
\end{equation}
complemented by the conditions (\ref{DW1}-\ref{DW2}). The solution of Eq.(\ref{LR0}), which obeys the boundary conditions (\ref{DW2}), is
\begin{equation}
	\delta \hat{f}_{lr}(x)={\Big \{}%
	\begin{array}{c}
		\hat{C}_{<}\cosh (\tilde{x}+\tilde{L})\text{, }-L<x<l \\
		\hat{C}_{>}\cosh (\tilde{x}-\tilde{L})\text{, }l<x<L\text{,}%
	\end{array}%
	\;.  \label{LR1}
\end{equation}
The matrices \(\hat{C}_{\lessgtr}\) are found from the matching conditions (\ref{DW1}-\ref{DW1a})
\begin{equation}
	\hat{C}_{\lessgtr }=\hat{C}_{\lessgtr }^{\text{(A)}}\cos (\varphi /2)+i%
	\mathit{\hat{X}}_{30}\hat{C}_{\lessgtr }^{\text{(B)}}\sin (\varphi /2)
	\label{LR2}
\end{equation}%
and find for $\hat{C}_{\lessgtr }^{\text{(A,B)}}$
\begin{eqnarray}
	\hat{C}_{\lessgtr }^{\text{(A)}} &\equiv &\hat{a}_{\lessgtr }=-4\frac{\kappa
		_{DW}}{\kappa _{\omega }}\frac{\cosh (\tilde{L}\mp \tilde{l})}{\sinh (2%
		\tilde{L})}A\mathit{\hat{X}}_{n2}\text{,}  \label{LR3a} \\
	\hat{C}_{\lessgtr }^{\text{(B)}} &\equiv &\hat{b}_{\lessgtr }=-4\frac{\kappa
		_{DW}}{\kappa _{\omega }}\frac{\cosh (\tilde{L}\mp \tilde{l})}{\sinh (2%
		\tilde{L})}B\mathit{\hat{X}}_{n2}  \label{LR3b}
\end{eqnarray}
where the signs \(\pm \) correspond to \(x\gtrless l\) and \(n=1\) for \(y\)
-chirality and \(n=2\) for \(x\)-chirality.

Having known the long-range Green's function \(\hat{f}_{lr}=\hat{f}_{lr,0}+\delta \hat{f}_{lr}\), we can find a change of the current in the
presence of a DW.

\subsection{Change of the Currents due to domain wall}

The corrections to the currents are
\begin{eqnarray}
	\delta I_{Q} &=&\left(\sigma _{F}/e\right)2\pi T\sum_{\omega \geqslant 0}\delta
	I_{Q,\omega }  \label{C1} \\
	\delta I_{sp} &=&\mu _{B}(\sigma _{F}/e^{2})2\pi T\sum_{\omega \geqslant
		0}\delta I_{sp,\omega }  \label{C1a}
\end{eqnarray}
and the partial currents \(\delta I_{Q,\omega}\) and \(
\delta I_{sp,\omega }\) are given by
\begin{eqnarray}
	\delta I_{Q,\omega } &=&i\{\delta \hat{f}_{lr}\partial _{x}\hat{f}_{lr,0}+%
	\hat{f}_{lr,0}\partial _{x}\delta \hat{f}_{lr}\}_{30}\text{,}  \label{C2b} \\
	\delta I_{sp,\omega } &=&i\{\delta \hat{f}_{lr}\partial _{x}\hat{f}_{lr,0}+%
	\hat{f}_{lr,0}\partial _{x}\delta \hat{f}_{lr}\}_{03}  \label{C2c}
\end{eqnarray}
We find
\begin{eqnarray}
	\delta I_{Q,\omega } &=&\kappa _{\omega }\{(\hat{C}^{\text{(B)}}+\hat{S}^{%
		\text{(B)}})\cdot \hat{a}-(\hat{C}^{\text{(A)}}+\hat{S}^{\text{(A)}})\cdot
	\hat{b}\}_{00}  \nonumber \\
	\label{C3} \\
	\delta I_{sp,\omega } &=&\kappa _{\omega }\{(\hat{C}^{\text{(B)}}+\hat{S}^{%
		\text{(B)}})\cdot \hat{a}-(\hat{C}^{\text{(A)}}+\hat{S}^{\text{(A)}})\cdot
	\hat{b}\}_{33} \nonumber \\
	\label{C3a}
\end{eqnarray}
Here, the matrices \(\hat{C}^{\text{(A,B)}}\) and \(\hat{S}^{\text{(A,B)}}\) are presented in the Appendix C (Eqs.(\ref{Ac1}-\ref{Ac1c})), and the matrices \(\hat{a}\equiv \hat{a}_{>}\), \(\hat{b}\equiv \hat{b}_{>}\) are defined in Eqs.(\ref{LR3a}-\ref{LR3b}).

Then, we find for the currents of Cooper pairs injected from the right and
left S\(_{m}\) reservoirs with equal chiralities and arbitrary spin polarizations
\begin{eqnarray}
	\delta I_{Q} &=&\delta I_{Q,\omega }\sin \varphi \text{,}  \label{C5} \\
	\delta I_{sp} &=&\delta I_{sp,\omega }\sin \varphi \text{. \ }  \label{C5a}
\end{eqnarray}
The critical currents \(\delta I_{Q,\omega}\) and \(\delta I_{sp,\omega}\) depend on the chiralities and polarizations of Cooper pairs propagating from the right and from the left. For the case {\it 
a) (\(xx\)) - chiralities, \(P\)-case (\(s=s_{r}=s_{l}\))} 
\begin{eqnarray}
	\delta I_{Q,\omega \text{;}P}^{(xx)} &=&-2\frac{\kappa_{DW}^{2}}{K_{0}}\qty(\frac{\kappa_{b}}{\kappa_{\omega}})^{2}F_{S-}^{2}%
	\frac{\cosh (\tilde{L}+\tilde{l})\cosh (\tilde{L}-\tilde{l})}{\sinh ^{2}(2%
		\tilde{L})}\text{ },  \label{C6a} \\
	\delta I_{sp,\omega \text{;}P}^{(xx)} &=&s\delta I_{Q,\omega \text{;}%
		P}^{(xx)}\text{. \ }  \label{C6b}
\end{eqnarray}
{\it (b) (\(xx\))} - chiralities, \(AP\)-case (\(s\equiv s_{r}=-s_{l}\)) we find
\begin{equation}
	\delta I_{Q,\omega \text{;}AP}^{(xx)}=-\text{\ }\delta I_{Q,\omega \text{;}%
		P}^{(xx)}\text{, }\delta I_{sp,AP}^{(xx)}=-\delta I_{sp,P}^{(xx)}
	\label{C7c}
\end{equation}
Comparing this equation and Eqs.(\ref{C6a}-\ref{C6b}), we see that the signs of the currents \(\delta I_{Q,AP}^{(xx)}\) and \(\delta I_{sp,AP}^{(xx)}\) are changed. 
\begin{figure}[ht!]
	\includegraphics[width=\columnwidth]{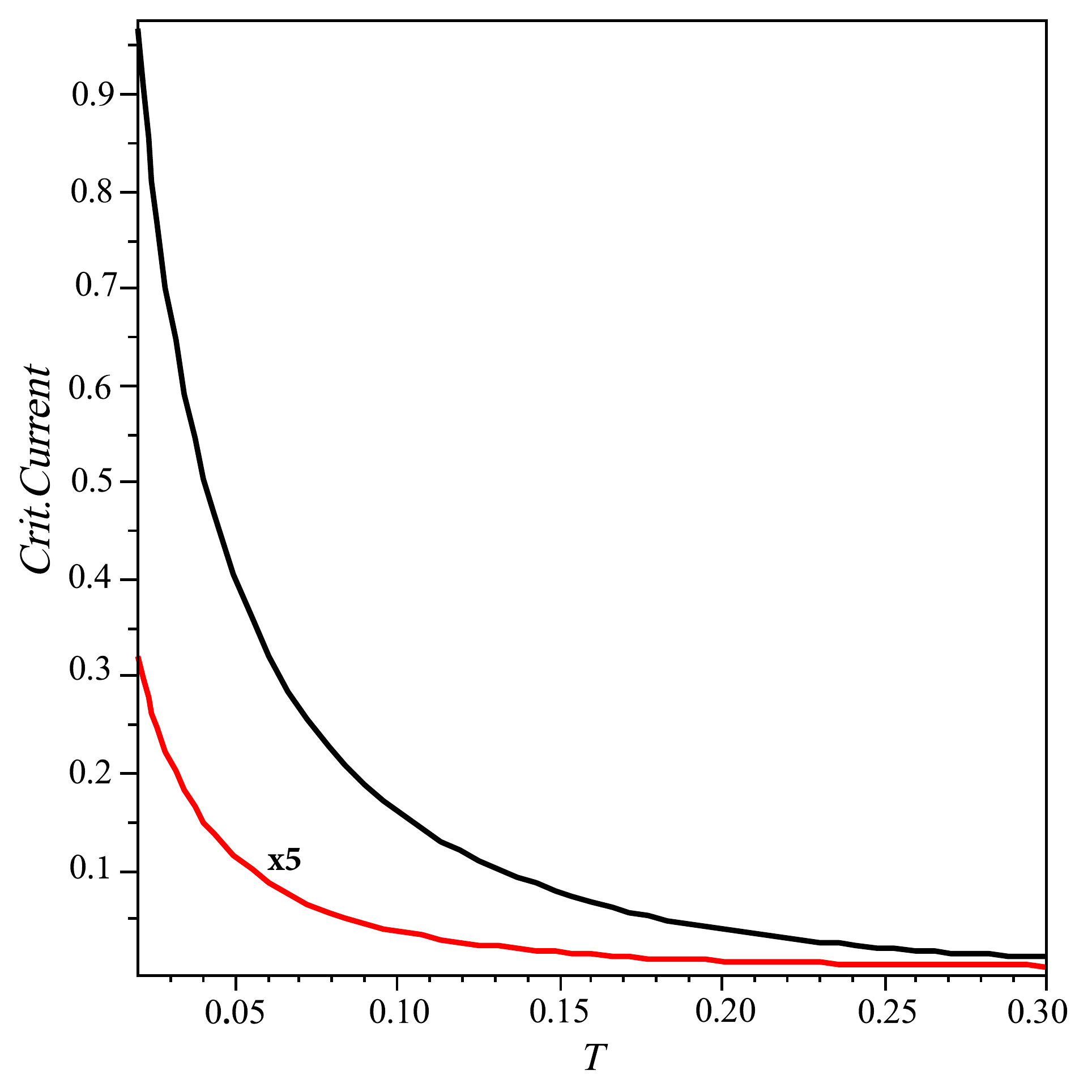} \vspace{-1mm}
	\vspace{-1mm}
	\caption{(Color online.) Temperature dependence of the normalized critical current $I_{c,0}$ in the absence of DW (black curve) and a change of the normalized critical current $\protect\delta I_{c}$ due to DW (red curve), which is substracted from $I_{c,0}$ to obtain the total current. For the sake of the presentation the magnitude of the latter is multiplied by the factor of 5. The change $\protect\delta I_{c}$ decreases the Josephson critical current $I_{c}$ if $I_{c,0}$ is not zero
and makes $I_{c}$ finite if $I_{c,0}=0$ (antiparallel spin orientations of triplet Cooper pairs injected from
the left and from the right).
The temperature $T$ and the exchange $J_{m}$ energy
in F $_{m}$ are normalized to $\Delta (0) $. The parameter $J\equiv
J_{m}/\Delta (0)$ is chosen to be equal to 3.(see Appendix D)}
	\label{Fig.3}
\end{figure}
Finally, in the case {\it c) (\(yy\)) - chiralities, \(P\)(\(AP\))-cases}, the currents are
\begin{eqnarray}
	\delta I_{Q,P}^{(yy)} &=&\delta I_{Q,P}^{(xx)}=\delta I_{Q,AP}^{(yy)}
	\label{C8} \\
	\delta I_{sp,P}^{(yy)} &=&-\delta I_{sp,P}^{(xx)}=\delta I_{sp,AP}^{(yy)}
	\label{C8b}
\end{eqnarray}
\begin{figure}[ht!]
	\includegraphics[width=\columnwidth]{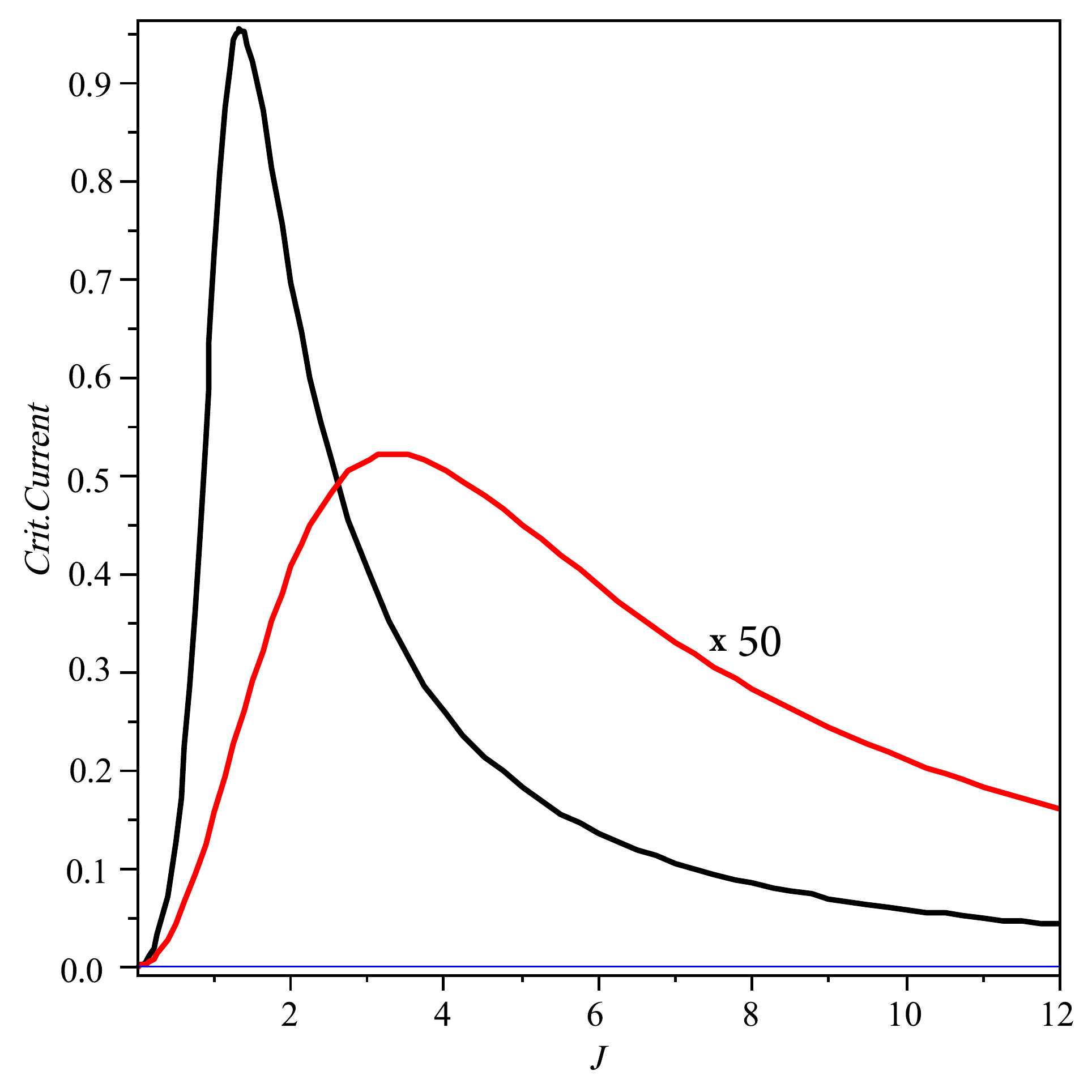} \vspace{-1mm}
	\vspace{-1mm}
	\caption{(Color online.) The dependence of the critical current $I_{c,0}(J_{m})$ on the exchange
field in F$_{m}$ film $J_{m}$ for two values of the normalized temperature and $L/\xi _{\Delta }=1$:
$T/\Delta (0)=0.05$ (black) and $T/\Delta (0)=0.3$ (red) (in the latter case, we multiplied 
this dependence by the factor 50 because the critical current decreases drastically with increasing $T$).}
	\label{Fig.4}
\end{figure}
In the case of the \(yy\)- chirality, the coefficients \(A^{(y,y)}\) and \(
B^{(y,y)}\) do not depend on the polarization \(s\). That is, the currents are equal for different spin orientations: \(\delta I_{Q,P}^{(yy)}=\delta I_{Q,AP}^{(yy)}\) and \(\delta I_{sp,P}^{(yy)}=\delta I_{sp,AP}^{(yy)}\).

The analysis of the obtained results shows that the DW reduces the Josephson charge and spin currents if Cooper pairs  injected from the right and left superconductors have parallel spin orientation. Thus, the action of the DW on the critical current in this case is analogous to the action of paramagnetic impurities, which decrease the penetration length of the LRTC \cite{bergeret_odd_2005,Fominov_Minigap_2006}.  In the case of antiparallel orientations, the DW makes the Josephson critical current finite. It is interesting to note that the maximum magnitude of the total Josephson current \(I_{Q}=|I_{Q,0}+\delta I_{Q}|\) is achieved at \(l=0\). This means that the Josephson energy has a minimum if the DW is located in the center of the junction for the case of parallel spin polarized Cooper pairs. Note also that the correction to the current $\delta I_{\omega ,0}$
is proportional to the square of $\kappa _{DW}$: $\delta I_{\omega ,0}\sim
K_{0}r_{\omega }^{2}=(\kappa _{DW}\kappa _{b})^{2}/(K_{0}\kappa _{\omega
}^{2})$. Thus, the contribution to the current due to DW does not depend on
whether the magnetisation vector $\mathbf{M}$ in the Bloch DW rotates clockwise or counterclockwise. The results of the change of the Josephson currents due to a single domain wall are also summarized in Table I.

In Fig.2 we plot the temperature dependence of the Josephson critical current \(I_{Q,0}(T)\) in the absence of a DW and a correction to the current due to a DW located in the center of the F film (\(l=0\)), see also Appendix D for details of the numerics. One can see that the critical current \(I_{Q,0}(T)\) and the correction due to a DW decay monotonously with increasing the temperature. For completeness  we show in Fig.3 the dependence \(I_{Q,0}(J)\) for two temperatures. A similar dependence shows the correction to the Josephson current due to a DW.

\subsection{Change of the Currents due to two domain walls}

We assume that the spacing between the nearest DWs is much larger than $\kappa _{J}$. In this case each DW contributes to the Josephson current
independently from others. Thus, the correction to the Josephson current due
to, for example two DWs, is given by
\begin{widetext}

\begin{equation}
\delta I_{Q,\omega }^{(xx)}=-2\frac{(\kappa _{b}\kappa _{DW}F_{S-})^{2}}{%
K_{0}\kappa _{\omega }^{2}\sinh ^{2}(2\tilde{L})}[\cosh (2\tilde{L})+\cosh (2%
\bar{l})+\cosh (2\delta l)]\sin \varphi \text{,}  \label{B}
\end{equation}%

\end{widetext}
where $\bar{l}=(\tilde{l}_{1}+\tilde{l}_{2})/2$, $\delta \tilde{l}=$ $\tilde{%
l}_{1}-\tilde{l}_{2}$. According to the assumption above $(\delta \tilde{l}%
\kappa _{J})\gg 1$. This formula means that the DW reduces the critical
current $I_{c}=I_{c0}+\delta I_{c}$ in the $P-$case and makes it finite in
the $AP-$case. The decrease of the current $I_{c}=I_{c0}-|\delta I_{c}|$ in
the $P-$case would be minimal if $\bar{l}=0$, i. e., the two DWs are located
in the center of the F film.

\section{Conclusions}

We have calculated the Josephson charge \(I_{Q}\) and spin \(I_{sp}\) currents in an S/F\(_{m}\)/Fl/F/Fl/F\(_{m}\)/S Josephson junction when the Josepson coupling is realized via different types of a long-range triplet component (LRTC).  The superconducting condensate in a thin magnetic layer F\(_{m}\) consists of singlet and triplet Cooper pairs penetrating from the S banks into the F\(_{m}\) film. The spin filters Fl pass only the triplet Cooper pairs which are long range in\ F because the magnetisation vector \(\mathbf{m}\) in F\(_{m}\) is perpendicular to the magnetisation vector \(\mathbf{M}||\mathbf{e}_{z}\) in the F film. The long-range triplet Cooper pairs,
penetrating into the F film, differ in chiralities, i. e., by orientation of
the vector \(\mathbf{m}\) ( \(\mathbf{m}||\mathbf{e}_{x}\) or \(\mathbf{m}||\mathbf{e}_{y}\)), and in polarization of the total spin of the triplet
Cooper pairs. First, we considered the case of a uniform magnetisation in F,
\(\mathbf{M}(x)=const\) , and of the absence of spin filters. Then, the
LRTC consists of equal numbers of fully polarized triplet pairs with opposite directions of the total spin \(s_{l,r}\) (the nematic case in terminology of Ref.\cite{Moor_nematic_2015}). In this case, the spin current \(I_{sp}^{(xx)}=I_{sp}^{(yy)}\) is zero and the Josephson current \(I_{Q}^{(xx)}=I_{Q}^{(yy)}\) is finite.

In the presence of the spin filters, both currents, \(I_{Q}^{(xx)}=I_{Q}^{(yy)}=\tilde{I}_{Q}(1+s_{r}s_{l})\sin \varphi \) and \(
I_{sp}^{(xx)}=I_{sp}^{(yy)}=\tilde{I}_{sp}(s_{r}+s_{l})\sin \varphi \), are
finite.\ If the chiralities and spin directions of the LRTC are equal, the
currents are finite and differ only by a prefactor. In the case of antiparallel spin orientations and \(s_{r}=-s_{l}=s\), the both currents are zero. If the triplet Cooper pairs injected on the right and on the left have different chiralities, spontaneous currents may arise: \(I_{Q}^{(xy)}=-\tilde{I}_{Q}(s_{r}+s_{l})\cos \varphi \), \(I_{sp}^{(xy)}=-\tilde{I}_{sp}(1+s_{r}s_{l})\cos \varphi \). This means that the currents may occur in the absence of the phase difference and the direction of the charge current depends on spins. The spontaneous currents may be the
reason for the Josephson diode effect, discussed recently\cite{Pal_2021}. All these results are summarized in Table I. 

We have studied the change of the charge and spin currents in the presence
of arbitrary number of DWs in the F film. It turns out that a DW reduces the
critical Josephson current if the spin directions of the Cooper pairs
injected from the right and left superconductors S\(_{m}\) are parallel (\(s_{r}=s_{l}\)). The critical current reaches a maximum if the DW is located
in the center of the F film. In the case of an antiparallel spins, \(
s_{r}=-s_{l}\), the critical current \(I_{c,0}^{\text{(AP)}}\) in the absence
of a DW is zero, but becomes finite in the presence of a DW.

The case of two DWs, which may be considered as a model of a skyrmion, is particularly interesting. The dependence of the change of the critical current \(\delta
I_{Q,\omega }^{(xx)}\) due to two DWs is given by Eq.(\ref{B}). In the case
of parallel spins (\(s_{r}=s_{l}\)), the critical current \(I_{Q,c}\) has a
maximum if the DWs are located in the center of the F film. In the case of
antiparallel spins (\(s_{r}=-s_{l}\)), the maximum \(I_{Q,c}\) corresponds to
the location of two DWs at the edges of the F film.

\section{Acknowledgements}

The authors acknowledge support from the Deutsche Forschungsgemeinschaft Priority Program SPP2137, Skyrmionics, under Grant No. ER 463/10.

\bibliography{literature} 

	\onecolumngrid
	\appendix


\section{Green's Functions $\hat{G}_{ik}$}

	\setcounter{equation}{0}
	\setcounter{figure}{0}
	\setcounter{table}{0}
	\makeatletter
	\renewcommand{\theequation}{A\arabic{equation}}
	\renewcommand{\thefigure}{A\arabic{figure}}

First we calculate the exact Green's functions \(\hat{G}_{ik}\) and show that they and the quasiclassical Green's functions \(\hat{g}^{(x)}(s)\) and \(\hat{g}^{(y)}(s)\) describe the fully polarized triplet Cooper pairs with spin \(s=\pm 1\). We use the Nambu indices defined in Ref.\cite{bergeret_odd_2005} , so that \(c_{n,s}=c_{s}\) and \(c_{n,s}=c^{\dagger}_{\bar{s}}\) for \(n=2\); \(c_{s}=c_{\uparrow}\) for \(s=1\) (\(\bar{s}=2\)) and \(c_{s}=c_{\downarrow}\) for \(s=2\) (\(\bar{s}=1\)). The Green's function \(\hat{G}_{12}\) is
\begin{align}
	\hat{G}_{12}(t,t')=&-i\expval{c_{ns}(t)\mathit{\hat{X}}_{12}c_{n's'}^{\dagger}(t*)}\notag\\
	=&-i\expval{c_{ns}(t)\hat{\tau}_{1}\otimes\hat{\sigma}_{2}c_{n's'}^{\dagger}(t*)}\notag\\
	=&-i\expval{c_{s}(t)\hat{\sigma}_{2}c_{\bar{s}'}^{\dagger}(t')+c_{\bar{s}}^{\dagger}(t)\hat{\sigma}_{2}c_{s'}(t')}\notag\\
	=&-i\expval{-c_{\uparrow}(z)c_{\uparrow}(t')+c_{\downarrow}(z)c_{\downarrow}(t')-c_{\uparrow}^{\dagger}(z)c_{\uparrow}^{\dagger}(t')+c_{\downarrow}^{\dagger}(z)c_{\downarrow}^{\dagger}(t')}\label{Eq: Green 12}
\end{align}
and
\begin{align}
	\hat{G}_{21}(t,t')=&-i\expval{c_{ns}(t)\mathit{\hat{X}}_{21}c_{n's'}^{\dagger}(t*)}\notag\\
	=&-i\expval{c_{\uparrow}(z)c_{\uparrow}(t')+c_{\downarrow}(z)c_{\downarrow}(t')-c_{\uparrow}^{\dagger}(z)c_{\uparrow}^{\dagger}(t')-c_{\downarrow}^{\dagger}(z)c_{\downarrow}^{\dagger}(t')}
\end{align}
Analogously, we obtain for \(\hat{G}_{11}\) and \(\hat{G}_{22}\)
\begin{align}
	\hat{G}_{11}(t,t')=&-i\expval{c_{\uparrow}(z)c_{\uparrow}(t')+c_{\downarrow}(z)c_{\downarrow}(t')+c_{\uparrow}^{\dagger}(z)c_{\uparrow}^{\dagger}(t')+c_{\downarrow}^{\dagger}(z)c_{\downarrow}^{\dagger}(t')}\\
	\hat{G}_{22}(t,t')=&-i\expval{c_{\uparrow}(z)c_{\uparrow}(t')-c_{\downarrow}(z)c_{\downarrow}(t')-c_{\uparrow}^{\dagger}(z)c_{\uparrow}^{\dagger}(t')-c_{\downarrow}^{\dagger}(z)c_{\downarrow}^{\dagger}(t')}\label{Eq: Green 22}
\end{align}
Combining Eqs.(\ref{Eq: Green 12}-\ref{Eq: Green 22}), one can write
\begin{align}
	\hat{G}^{(x)}(t,t')\equiv&\hat{G}_{11}-s\hat{G}_{22}=-2i\expval{c_{s}(t)c_{s}(t')+c_{s}^{\dagger}(t)c_{s}^{\dagger}(t')}\label{Eq: Gx}\\
	\hat{G}^{(y)}(t,t')\equiv&\hat{G}_{12}+s\hat{G}_{21}=2i(-1)^{s}\expval{c_{s}(t)c_{s}(t')+c_{s}^{\dagger}(t)c_{s}^{\dagger}(t')} \label{Eq.: Gy}
\end{align}
Eqs.(\ref{Eq: Gx},\ref{Eq.: Gy}) show that both Green's functions \(\hat{G}^{(x)}\) and \(\hat{G}^{(y)}\) are off-diagonal in the Nambu-space and define triplet Cooper pairs with spin up \((s=1)\) and down \((s=-1)\), which describe a fully polarized triplet component. Since the matrix structure of the Green's functions does not change upon going over to the quasiclassical functions,
\begin{equation}
	\hat{g}_{BVE}=\frac{i}{\pi}\nu\int\dd{\xi}\hat{G}
\end{equation}
the same statement is true for matrix functions \(\hat{g}_{BVE}\).\\
Note the transformation suggested by Ivanov-Fominov
\begin{align}
	\hat{g}=&\hat{U}\cdot\hat{g}_{BVE}\cdot\hat{U}^{\dagger}\\
	\hat{U}=&\frac{1}{2}\qty(\mathit{\hat{X}}_{00}+i\mathit{\hat{X}}_{33})\cdot\qty(\mathit{\hat{X}}_{00}-i\mathit{\hat{X}}_{33})
\end{align}
transform the function \(\hat{g}_{BVE}\) introduced in Ref.\cite{bergeret_odd_2005} into the functions \(\hat{g}\), employed here. It does not change the relations because the matrix \(\hat{U}\) commutes with the matrices \(\mathit{\hat{X}}_{00}\) and \(\mathit{\hat{X}}_{33}\).\\
These functions arise as a result of the action of the spin filters (we set \(T=1\), \(U=\pm1\)).
\begin{align}
	\hat{G}^{(x)}=&T_{33}\cdot\hat{G}_{11}\cdot T_{33}^{\dagger}\\
	\hat{G}^{(y)}=&T_{33}\cdot\hat{G}_{12}\cdot T_{33}^{\dagger}\\
	T_{33}=&\frac{1}{\sqrt{2}}\qty(T+U\mathit{\hat{X}}_{33})
\end{align}
\section{Charge and Spin currents}
	\setcounter{equation}{0}
	\setcounter{figure}{0}
	\setcounter{table}{0}
	\makeatletter
	\renewcommand{\theequation}{B\arabic{equation}}
	\renewcommand{\thefigure}{B\arabic{figure}}

The charge density \(\rho\) is equal to
\begin{align}
	\rho(r,t')=&C_{Q}\sum_{p}\expval{c_{ns}^{\dagger}(t)\mathit{\hat{X}}_{30}c_{n's'}(t')}\notag\\
	=&C_{Q}\sum_{p}\expval{c_{ns}^{\dagger}(t)\hat{\tau}_{3}c_{n's'}(t')}=\expval{c_{s}^{\dagger}(t)c_{s}(t')-c_{\bar{s}}(t)c_{\bar{s}}^{\dagger}(t')}\notag\\
	=&C_{Q}\sum_{p}\expval{c_{\uparrow}^{\dagger}(t)c_{\uparrow}(t')+c_{\downarrow}^{\dagger}(t)c_{\downarrow}(t')-c_{\uparrow}(t)c_{\uparrow}^{\dagger}(t')-c_{\downarrow}(t)c_{\downarrow}^{\dagger}(t')}
\end{align}
where \(C_{Q}\) is a constant which will be defined below. The operators \(c_{ns}^{\dagger}(t')\),  \(c_{n's'}(t)\), as before, depend on times \(t,t'\).
For equals times \(t=t'\), we obtain
\begin{align}
	\rho(t)=&2C_{Q}\sum_{p}\expval{c_{\uparrow}^{\dagger}c_{\uparrow}+c_{\downarrow}^{\dagger}c_{\downarrow}}\notag\\
	=&-2iC_{Q}\sum_{p}\qty{\hat{G}}_{30}=\frac{2}{\pi}C_{Q}\nu(0)\qty{\hat{g}_{BVE}}_{00}
\end{align}
Here, \(\hat{g}_{BVE}\) is the quasiclassical Green's function derived in \cite{bergeret_odd_2005}. The magnetic moment is
\begin{align}
	M=&C_{M}\sum_{p}\expval{c_{ns}^{\dagger}(t)\mathit{\hat{X}}_{03}c_{n's'}(t')}\notag\\
	=&C_{M}\sum_{p}\expval{c_{s}^{\dagger}\sigma_{3}c_{s'}+c_{\bar{s}}\hat{\sigma}_{3}c_{\bar{s}'}^{\dagger}}\notag\\
	=&C_{M}\sum_{p}\expval{c_{\uparrow}^{\dagger}(t)c_{\uparrow}(t')+c_{\downarrow}^{\dagger}(t)c_{\downarrow}(t')}
\end{align}
For equal times \(t=t'\), we obtain
\begin{align}
	M(t)=&2C_{M}\sum_{p}\expval{c_{\uparrow}^{\dagger}(t)c_{\uparrow}(t)+c_{\downarrow}^{\dagger}(t)c_{\downarrow}(t)}\notag\\
	\Rightarrow&2iC_{M}\sum_{p}\qty{\hat{G}}_{03}\Rightarrow\frac{2}{\pi}C_{M}\qty{\hat{g}_{BVE}}_{33}
\end{align}
To find the formula for the charge (spin) current, consider the Usadel equation for the Keldysh function
\begin{equation}
	\hat{\tau}_{3}\cdot\partial_{t}\hat{g}+\partial_{t*}\hat{g}\cdot\tau_{3}=D_{\text{F}}\partial_{x}(\hat{g}\cdot\partial_{x}\hat{g})+iJ\qty[\mathit{\hat{X}}_{33},\hat{g}]\label{Eq: Usd}
\end{equation}
Introducing \(\bar{t}=(t+t')/2\) and \(\tau=t-t'\), Eq.(\ref{Eq: Usd}) can be written as
\begin{equation}
	\frac{1}{2}\partial_{\bar{t}}\qty[\hat{\tau}_{3},\hat{g}]_{+}+\partial_{\tau}\qty[\hat{\tau}_{3},\hat{g}]=D_{F}\partial_{x}(\hat{g}\cdot\partial_{x}\hat{g})+iJ\qty[\mathit{\hat{X}}_{33},\hat{g}]\label{EQ: Usd traf}
\end{equation}
We multiply Eq.(\ref{EQ: Usd traf}) first by \(\mathit{\hat{X}}_{30}\), then by \(\mathit{\hat{X}}_{03}\) and calculate the trace. We get the law of conservation of the charge and the magnetization
\begin{equation}
	\pdv{\rho}{\bar{t}}=-\partial_{x}j_{Q} \qquad
	\pdv{M}{\bar{t}}=-\partial j_{sp}
\end{equation}
where the charge current \(j_{Q}\) is equal to
\begin{align}
	j_{Q}=&-\frac{\sigma_{n}}{e}2\pi T \sum_{\omega\geq0}\frac{1}{4}\Tr{\hat{\tau}_{3}\hat{g}\nabla\hat{g}}\notag\\
	=&-\frac{\sigma_{n}}{e}2\pi T \sum_{\omega\geq0}\qty{\hat{g}\nabla\hat{g}}_{30}
\end{align}
and the spin current \(j_{sp}\) is given by
\begin{equation}
	\vb{j}_{sp}=-\mu_{B}\frac{\sigma_{n}}{e^{2}}(i\pi T) \sum_{\omega\geq0}\nabla\qty{\hat{g}}_{03}
\end{equation}
The charge density \(\rho\) is
\begin{equation}
	\rho=e\nu(0)(i2\pi T)\sum_{\omega\geq0}\qty{\hat{g}}_{00}
\end{equation}
The Drude conductivity \(\sigma_{n}\) is
\begin{equation}
	\sigma_{n}=2\nu(0)D_{n}e^{2}
\end{equation}
The magnetic moment \(M_{z}\) is (see, for example, \cite{bergeret_odd_2005}, Eq.(A28))
\begin{equation}
	M_{z}=\mu_{B}\nu(0)(i2\pi T)\sum_{\omega\geq0}\qty{\hat{g}}_{33}
\end{equation}

\section{Coefficients in the Change of the Currents due to two DWs}
	\setcounter{equation}{0}
	\setcounter{figure}{0}
	\setcounter{table}{0}
	\makeatletter
	\renewcommand{\theequation}{C\arabic{equation}}
	\renewcommand{\thefigure}{C\arabic{figure}}

The coefficients \(\hat{C}^{\text{(A,B)}}\) and \(\hat{S}^{\text{(A,B)}}\) in
Eqs.(\ref{C3}-\ref{C3a}) are determined by Eqs.(\ref{5a}-\ref{5b}). They are
equal to
\begin{eqnarray}
\hat{C}^{\text{(A)}} &=&\frac{\kappa _{b}}{2\kappa _{\omega }}(\mathit{\hat{X%
	}}_{r}+\mathit{\hat{X}}_{l})\cos (\varphi /2)F_{S-}\text{,}  \label{Ac1} \\
	\hat{C}^{\text{(B)}} &=&\frac{\kappa _{b}}{2\kappa _{\omega }}(\mathit{\hat{X%
		}}_{r}-\mathit{\hat{X}}_{l})\sin (\varphi /2)]F_{S-}\text{,}  \label{Ac1a} \\
		\hat{S}^{\text{(A)}} &=&\frac{\kappa _{b}}{2\kappa _{\omega }}[(\mathit{\hat{%
				X}}_{r}-\mathit{\hat{X}}_{l})\cos (\varphi /2)F_{S-}\text{,}  \label{Ac1b} \\
		\hat{S}^{\text{(B)}} &=&\frac{\kappa _{b}}{2\kappa _{\omega }}(\mathit{\hat{X%
			}}_{r}+\mathit{\hat{X}}_{l})\sin (\varphi /2)F_{S-}  \label{Ac1c}
\end{eqnarray}
			
The matrices \(\hat{a}\), \(\hat{b}\) equal
			
\begin{eqnarray}
			\hat{a} &=&-4r_{\omega }\frac{\cosh (\tilde{L}\pm \tilde{l})}{\sinh (2\tilde{%
					L})}A\cos (\varphi /2)\mathit{\hat{X}}_{n2}\text{; }  \label{Ac2} \\
			\hat{b} &=&-4r_{\omega }\frac{\cosh (\tilde{L}\pm \tilde{l})}{\sinh (2\tilde{%
					L})}B\sin (\varphi /2)\mathit{\hat{X}}_{n2}\text{.}  \label{Ac2a}
\end{eqnarray}
			
with \(n=1,2\) for \(y\)- and \(x\)-chiralities.

\section{Details of the numerics}
	\setcounter{equation}{0}
	\setcounter{figure}{0}
	\setcounter{table}{0}
	\makeatletter
	\renewcommand{\theequation}{D\arabic{equation}}
	\renewcommand{\thefigure}{D\arabic{figure}}
The critical current \(\tilde{I}_{0}(t)\) of the considered Josephson junction without DWs is
\begin{eqnarray}
	\tilde{I}_{0}(t) &=&I_{0}N(t)\text{, ,}  \label{N1} \\
	I_{0}&=&\frac{\sigma_{\text{F}}\Delta}{e}\xi_{\Delta}\kappa_{b}^{2}\\
	N(t) &=&2\pi t\sum_{n\geqslant 0}\qty[\text{Im}\frac{\tilde{\Delta}(t)}{\sqrt{%
	(t_{n}+i\tilde{J}_{m})^{2}+\tilde{\Delta}(t)^{2}}}]^{2}\frac{1}{\sqrt{t_{n}}
	\sinh (2\tilde{L}\sqrt{t_{n}})}  \label{N1a}
\end{eqnarray}
where \(\xi _{\Delta }=\sqrt{D_{F}/2\Delta }\), \(\tilde{L}=L/\xi
_{\Delta }\), \(t_{n}=\pi t(2n+1)\), \(t=T/\Delta(0) \).\\
The correction to the current due to a single DW is

\begin{eqnarray}
\delta \tilde{I}(t) &=&-I_{DW}N_{DW}(t)\text{, }  \label{N2} \\
I_{DW} &=&\frac{\sigma _{F}\Delta }{e}(\xi _{\Delta }\kappa _{DW}\kappa
_{b})^{2}\xi _{\Delta }\text{, } \\
N_{DW}(t) &=&2\pi t\sum_{n\geqslant 0}[\text{Im}\frac{\tilde{\Delta}(t)}{%
\sqrt{(t_{n}+i\tilde{J}_{m})^{2}+\tilde{\Delta}(t)^{2}}}]^{2}\frac{\cosh (%
\sqrt{t_{n}}(\tilde{L}+\tilde{l}))\cosh ((\sqrt{t_{n}}(\tilde{L}-\tilde{l})))%
}{t_{n}(\sinh (2\tilde{L}\sqrt{t_{n}}))^{2}}
\end{eqnarray}%

The temperature dependence of \(\tilde{\Delta}(t)\equiv \Delta (T)/\Delta(0)\) can be approximated as
\begin{equation}
\Delta (T) \cong \Delta(0)\tanh (1.74\sqrt{(T_{c}/T-1)})
\label{N2new} 
\end{equation}
In the limits of \(T=0\) and \(T\Rightarrow T_{c}\) it reproduces the limiting expressions (see, for example \cite{abrikosov_fundamentals_1988})
\begin{align}
	2\Delta(0)\cong3.5T_{c}\\
	\Delta(T)|_{T\Rightarrow T_{c}}\cong3.06\sqrt{(T_c-T)T_{c}}
\end{align}

\end{document}